\documentclass[preprint,aps,prl,showpacs,amsmath,amssymb]{revtex4}
\usepackage{graphics}
\usepackage{epsf,graphicx,pstricks,epsfig,psfrag}
\usepackage{amsmath}
\usepackage{amssymb}

\begin{document}

\title{Coexistence of superconductivity and charge-density waves
in a two-dimensional Holstein model at half-filling}
\author{S.~Sykora, A.~H\"{u}bsch and K.~W.~Becker}

\affiliation{
  Institut f\"{u}r Theoretische Physik,
  Technische Universit\"{a}t Dresden, D-01062 Dresden, Germany} 

{\today}


\begin{abstract}

In one dimension the coupling of electrons to phonons leads to a 
transition from a metallic to a Peierls distorted insulated state if the 
coupling exceeds a critical value. On the other hand, in two dimensions the
electron-phonon interaction may also lead to the formation of Cooper pairs. 
This competition of superconductivity and charge order (in conjunction with 
a lattice distortion) is studied in this letter by means of the 
projector-based renormalization method (PRM). Increasing the electron-phonon 
interaction, we find a crossover behavior between a purely superconducting 
state and a charge-density wave where a well-defined parameter range of 
coexistence of superconductivity and lattice distortion exists.

\end{abstract}

\maketitle


\textit{Introduction.}
In recent years, the interest in two-dimensional electron-phonon models has 
been considerably renewed, mainly triggered by strong experimental indications 
that the electron-phonon interaction has a substantial impact on the 
properties of high-$T_{c}$ cuprates \cite{EP_in_HTC}, and by the discovery of 
superconductivity in MgB$_{2}$ below a rather high $T_{c}$ of about 
39~K \cite{MgB2}.  Furthermore, in the context of quantum-phase transitions, 
such systems are also of general theoretical interest because the 
electron-phonon interaction causes a competition between a structural 
instability and superconductivity. As pointed out by many 
authors (e.g. \cite{A15}, \cite{PZ89}, \cite{TP95}, \cite{M90}),  this 
competition may play an important role in $A15$ materials, (Ba,K)BiO$_{3}$ 
or high-temperature superconductors.

The simplest general model of a coupled electron-phonon system on a square 
lattice is given by  
\begin{eqnarray}
\label{1}
\mathcal{H} &=&  {\cal H}_0 + {\cal H}_1 \\
{\cal H}_0 &=& \sum_{{\bf k},\sigma} \varepsilon_{\bf k} c_{{\bf
 k},\sigma}^{\dag} c_{{\bf k},\sigma} +
 \sum_{\bf q} \omega_{\bf q} b_{\bf q}^{\dag} b_{\bf q} \nonumber \\
 {\cal H}_1 &=&
\frac{1}{\sqrt{N}} \sum_{{\bf k},{\bf q},\sigma} g_{\bf q} 
\left\{ b_{\bf q}^{\dag} c_{{\bf k},\sigma}^{\dag}
 c_{{\bf k} + {\bf q},\sigma} + b_{\bf q} c_{{\bf k} + {\bf q},\sigma}^{\dag}
 c_{{\bf k},\sigma} \right\}. \nonumber 
\end{eqnarray}
Here, $c^{\dagger}_{\mathbf{k},\sigma}, c_{\mathbf{k},\sigma}$ and 
$b_{\mathbf k}^{\dagger}, b_{\mathbf k}$ are the creation and annihilation 
operators of electrons and phonons. Assuming an electron hopping between 
nearest-neighbor sites on a square lattice, the electronic dispersion is 
given by $\varepsilon_{\mathbf k}= -2t (\cos k_xa + \cos k_ya) - \mu$, where 
$\mu$ is the chemical potential. Moreover, $\omega_{\mathbf q}$ is the phonon 
energy, and $g_{\mathbf q}$ is the wave vector dependent coupling between the 
electrons and phonons. In the case of dispersion-less phonons $\omega_0$
and a q-independent coupling $g$, the model \eqref{1} reduces to the
Holstein model. Then, the last term describes a local interaction 
between a lattice displacement and the local electronic density. For a more 
realistic description of high-$T_{c}$ materials, we would have to add a 
Hubbard-U term to the model.

The interaction between a possible superconducting (SC) state and an 
insulating Peierls charge-density wave (CDW) phase in the 2d-Holstein model 
has been subject to a number of Monte Carlo (QMC) studies \cite{QMC}. 
However, their mutual influence is not yet fully understood, because only 
correlation functions, namely SC and CDW susceptibilities, can be studied 
within QMC calculations but not the fundamental order parameters themselves.
Nevertheless, it is clear that the two phases compete: the appearance of 
strong correlations of one kind suppresses the development of correlations 
of the other kind. But these QMC studies could not clarify whether a 
coexistence of the two phases is possible or not. 

On the other hand, particular attention received a work by Bilbro and McMillan 
\cite{BM} from 1976 where the competition of superconductivity and the 
martensitic transformation in $A 15$ compounds was studied by a mean-field 
treatment of an appropriate model Hamiltonian and where a possible coexistence 
of a Peierls gap and a SC gap for a common portion of the Fermi surface was 
proposed. The Bilbro-McMillan model has been the basis for further 
theoretical work till now, for instance in the context of high-temperature 
superconductors \cite{GSV03},\cite{GV07} and heavy fermion 
compounds \cite{JBYM07}. However, to our knowledge two open problems in the 
Bilbro-McMillan model remain to be solved: First, all fluctuations beyond 
mean-field theory are neglected so that the equations for SC and CDW gaps are 
given by weak coupling expressions and any influence of a possible 
renormalization of the phonon energies is not included. Second, the 
electron-phonon interaction should be part of the basic microscopic model to 
have access to phonon properties.

Here, we tackle these two open questions by means of the recently developed 
projector-based renormalization method (PRM) \cite{PRM}. In this way we are 
able to include fluctuations beyond mean-field theory and to take into account 
additional renormalization effects like phonon softening. Furthermore, we 
directly access the order parameters of the two phases where, in agreement 
with reference \cite{BM}, we find a parameter range of coexistence of SC and 
CDW at half-filling. Depending on the strength of the electron-phonon 
coupling the Peierls phase will be suppressed by the superconducting phase 
or vice versa.


\textit{Projector-based renormalization method.}
The PRM \cite{PRM} starts from the usual decomposition of a many-particle 
Hamiltonian into a solvable unperturbed part ${\cal H}_0$ and a 
perturbation ${\cal H}_1$ where ${\cal H}_1$ does not contain any part 
that commutes with ${\cal H}_0$. Thus, the perturbation $\mathcal{H}_{1}$ 
consists of transitions between the eigenstates of ${\cal H}_0$ with 
non-vanishing transition energies. The basic idea of the PRM is the construction of an effective 
Hamiltonian ${\cal H}_\lambda = {\cal H}_{0,\lambda} + {\cal H}_{1,\lambda}$
with renormalized parts ${\cal H}_{0,\lambda}$ and ${\cal H}_{1,\lambda}$
where all transitions with transition energies 
$|E_{0,\lambda}^n -E_{0,\lambda}^m|$ larger than a given cutoff energy 
$\lambda$ are eliminated. Here, $E_{0,\lambda}^n$ and $E_{0,\lambda}^m$ 
denote the eigenenergies of ${\cal H}_{0,\lambda}$. 

The renormalization procedure starts from the cutoff energy $\lambda=\Lambda$ 
of the original model ${\cal H}$ and proceeds in steps of $\Delta \lambda$ 
to lower values of $\lambda$. Every renormalization step is performed by 
means of an unitary transformation where the elimination 
between the cutoffs $\lambda$ and $(\lambda - \Delta \lambda)$
reads  
\begin{eqnarray}
  \label{3}
  \mathcal{H}_{(\lambda-\Delta\lambda)} &=& 
  e^{X_{\lambda,\Delta\lambda}} \, \mathcal{H}_{\lambda} \,
  e^{-X_{\lambda,\Delta\lambda}}.
\end{eqnarray}
Here, the generator $X_{\lambda,\Delta \lambda}$ of the unitary
transformation has to be fixed appropriately (for details see 
Ref.~\onlinecite{PRM}). In this way difference equations are derived which 
connect the parameters of ${\cal H}_\lambda$ with those of 
${\cal H}_{(\lambda - \Delta \lambda)}$, and which we call renormalization 
equations.

The limit $\lambda \rightarrow 0$ provides the desired effective Hamiltonian 
$ 
  \tilde{\cal H}= {\cal H}_{\lambda \rightarrow 0} 
  = {\cal H}_{0, \lambda \rightarrow 0}
$
where the elimination of the transitions originating from the perturbation 
$\mathcal{H}_{1}$ leads to a renormalization of the parameters of 
${\cal H}_{0,\lambda \rightarrow 0}$. Note that $\tilde{\cal H}$ 
is diagonal or at least quasi-diagonal and allows to evaluate
physical quantities. The final results depend on the parameter values of the 
original Hamiltonian $\mathcal{H}$. Finally, note that $\tilde{\cal H}$ and 
${\cal H} $ have the same eigenvalue problem because both Hamiltonians are 
connected by an unitary transformation.


\textit{SC and CDW phases in the half-filled Holstein model}
We now want to apply the PRM approach to the Holstein model 
at half-filling in order to study the interplay between SC and CDW phases. 
For this purpose a uniform description of both the SC and the insulating CDW 
phase has to be found. In the SC state the gauge symmetry is broken. 
Therefore, following Ref. \onlinecite{PRM_SC}, a field which breaks gauge 
symmetry should be added to the Hamiltonian. Similarly, for half-filling one 
has to take into account that the unit cell of the system can be doubled in 
the case of a dimerized insulating ground state \cite{PRM_Holstein}. 
Therefore, we add to the Hamiltonian symmetry breaking fields 
as follows
\begin{eqnarray}
\label{4}
&& \mathcal{H} \Rightarrow \mathcal{H} +  
   \sum_{\bf k}   \left(       \Delta^{\mathrm{s}}_{\bf k} 
c_{{\bf k},\uparrow}^{\dag}  c_{-{\bf k},\downarrow}^{\dag} + 
 {\Delta^{\mathrm{s}}_{\mathbf k}}^* 
c_{-{\bf  k},\downarrow} c_{{\bf k},\uparrow} \right)  \\
&& \hspace*{0.1cm} +
\frac{1}{2} \sum_{{\bf k},\sigma} \left( \Delta^{\mathrm{p}}_{\bf k} \,
c_{{\bf k},\sigma}^{\dag} c_{{\bf k}-{\bf Q},\sigma} +{\rm h.c} \right) 
+ \sqrt{N}\Delta^b (b_{\bf Q}^\dagger+ b_{\bf Q}) \nonumber
\end{eqnarray}
where all fields are assumed to be 
infinitesimally small ($\Delta^{\mathrm s}_{\bf k}\rightarrow 0$,
$\Delta^{\mathrm{p}}_{\bf k} \rightarrow 0, \Delta^b  \rightarrow 0$). 
${\bf Q}= (\pi/a,\pi/a)$ is the characteristic wave vector of the CDW phase.
Due to the doubling of the unit cell in the insulating
phase the Hamiltonian is best rewritten in the reduced Brillouin zone where
we have $\varepsilon_{{\bf k}- {\bf Q}} = -\varepsilon_{\bf k}$.
Exploiting in addition the coupling of the creation operator 
$c_{{\bf k}, \sigma}^\dagger$
to the annihilation operator 
$c_{{\bf -k}, -\sigma}$ due to superconductivity a 
four-dimensional compact vector notation for the electronic 
one-particle operators can be  introduced
\begin{eqnarray}
\label{5}
\underline{c}_{\bf k}^\dag   
&=& \left(
\begin{array}{llll}
c_{-{\bf k}+{\bf Q},\downarrow} & c_{{\bf k},\uparrow}^\dag & 
c_{-{\bf k},\downarrow} & c_{{\bf k}-{\bf Q},\uparrow}^\dag
\end{array}
\right)
\end{eqnarray}
The renormalized Hamiltonian, after all transitions 
with energies larger than $\lambda$ have been integrated out, can again be
divided into ${\cal H}_\lambda =  {\cal H}_{0,\lambda}+ {\cal H}_{1,\lambda}$.
For ${\cal H}_{0,\lambda}$ we make the ansatz  
\begin{eqnarray}
\label{6}
 \mathcal{H}_{0,\lambda} &=&  \sum_{{\bf k}\in {\rm r.BZ}} 
\underline{c}_{\bf k}^\dag   
\hat{H}^{\rm el}_{{\bf  k},\lambda} \underline{c}_{\bf k} \\
&+& \sum_{\alpha=0,1} \sum_{{\bf q} \in {\rm r.BZ}} 
  \omega_{\alpha ,{\bf q},\lambda} b_{\alpha,{\bf q}}^{\dag}
 b_{\alpha,{\bf q}} +E_\lambda \nonumber 
\end{eqnarray}
with 
\begin{equation*}
\hat{H}^{\rm el}_{{\bf k},\lambda}  = \left(
\begin{array}{cccc}
\varepsilon_{} & 0 & -\Delta^{\mathrm{p}}_{} 
& \Delta^{\mathrm{s}\, *} \\
0 & \varepsilon_{} & 
\Delta^{\mathrm{s}}_{} & \Delta^{\mathrm{p}}_{} \\
-\Delta^{\mathrm{p}}_{} & 
\Delta^{\mathrm{s}\, *}_{} 
& -\varepsilon_{} & 0 \\
\Delta^{\mathrm{s}}_{} & 
\Delta^{\mathrm{p}}_{} & 0 & 
-\varepsilon_{}
\end{array}
\right)_{{\bf k},\lambda}
\end{equation*}
where all parameters now depend on ${\bf k}$ and $\lambda$
due to renormalization processes.
Note that the symmetry breaking fields have been included in 
${\cal H}_{0,\lambda}$, and the 
phonon field term $\sim \Delta^b$ from \eqref{4} was incorporated 
in redefined phonon operators. The new operators 
$b_{0,\bf q}^{(\dagger)}= b_{{\bf q}- {\bf Q}}^{(\dagger)}$ and  
$b_{1,\bf q}^{(\dagger)}= b_{\bf q}^{(\dagger)}$
characterize the phonon branches in the reduced Brillouin zone  
with energies
$\omega_{0,{\bf q},\lambda} = \omega_{{\bf q} - {\bf Q},\lambda}$
and  $\omega_{1,{\bf q},\lambda} = \omega_{{\bf q},\lambda}$. 
The renormalized interaction ${\cal H}_{1,\lambda}$
can be compactly written as well
\begin{equation}
\label{7}
 \mathcal{H}_{1,\lambda} = \frac{1}{\sqrt{N}} \sum_{{\bf k},{\bf q} 
\in {\rm r.BZ}}
  \sum_{\gamma \in \{0,1\}} \left\{ b_{\gamma,{\bf q}}^{\dag} 
\underline{c}_{\bf
  k}^{\dag}\, \hat{H}_{\gamma,{\bf k},{\bf q},\lambda}^{\rm w} \,
\underline{c}_{{\bf k} + {\bf q}} + \mbox{h.c.} 
  \right\} 
\end{equation}
where the elements of the new $4\times 4$ matrix 
$ \hat{H}_{\gamma,{\bf k},{\bf q},\lambda}^{\rm w}$ again depend on  
$\lambda$ and on wave vectors ${\bf q}$ and ${\bf k}$. 
Note that the general structure of ${\cal H}_{0,\lambda}$ and 
${\cal H}_{1,\lambda}$ and thus of the matrices 
$\hat{H}^{\rm el}_{{\bf k},\lambda}$
and $\hat{H}_{\gamma,{\bf k},{\bf q},\lambda}^{\rm w}$
remains always the same during the renormalization procedure 
and agrees with that of the corresponding matrix of the 
general electron-phonon model \eqref{1}. The initial values of the 
$\lambda$-dependent parameters in \eqref{6}, \eqref{7} are fixed by 
the original Holstein model.

The eigenvalue problem of ${\cal H}_{0,\lambda}$ 
can be solved analytically. For this purpose we introduce new
$\lambda$ dependent operators $a_{\alpha,{\bf k}, \lambda}$, 
$(\alpha=1,\cdots,4)$: $\underline{a}_{{\bf k},\lambda} = \hat{D}_{{\bf
    k},\lambda}\, \underline{c}_{\bf k}$ where the four-dimensional vector 
notation is used. Here, we defined
\begin{equation}
\label{9}
\hat{D}_{{\bf
    k},\lambda} =
\left(
\begin{array}{cccc}
u & 0 & v^{\mathrm{p}} & -v^{\mathrm{s}} \\
0 & u & -v^{\mathrm{s}} & -v^{\mathrm{p}} \\
-v^{\mathrm{p}} & v^{\mathrm{s}} & u & 0 \\
v^{\mathrm{s}} & v^{\mathrm{p}} & 0 & u
\end{array}
\right)_{{\bf k},\lambda}  .
\end{equation}
The condition 
$
(u_{{\bf k},\lambda})^{2} + (v_{{\bf k},\lambda}^{\rm p})^{2} + 
(v_{{\bf k},\lambda}^{\rm s})^{2} = 1
$
guarantees that the usual anti-commutator relations for 
$a_{\alpha,{\bf k}, \lambda}$ are fulfilled,
$ 
\left[a_{\alpha,{\bf k},\lambda}^{\dag},a_{\alpha',{\bf
 k}',\lambda}\right]_{+} = \delta_{{\bf k},{\bf k}'}
 \delta_{\alpha,\alpha'}
$.
For the electronic part of ${\cal H}_{0,\lambda}$ one obtains
\begin{eqnarray}
\label{10}
\mathcal{H}_{0,\lambda}^{\mathrm{el}} &=& \sum_{{\bf k} \in {\rm r.BZ}} 
\left\{ 
  E_{1,{\bf k},\lambda} \left( 
    a_{1,{\bf k},\lambda}^{\dag} a_{1,{\bf k},\lambda} 
    + a_{2,{\bf k},\lambda}^{\dag} a_{2,{\bf k},\lambda} 
  \right)
\right.   \nonumber \\
&& \left.
  \, + E_{2,{\bf k},\lambda} 
  \left( 
    a_{3,{\bf k},\lambda}^{\dag} a_{3,{\bf  k},\lambda} 
    + a_{4,{\bf k},\lambda}^{\dag} a_{4,{\bf k},\lambda} 
  \right) 
\right\} 
\end{eqnarray}
where the energies are given by 
\begin{eqnarray}
\label{11}
E_{1/2,{\bf k},\lambda} &=& 
\pm 
\sqrt{ 
      \varepsilon_{{\bf k},\lambda}^{2} 
  + |\Delta_{{\bf k},\lambda}^{\mathrm{p}}|^{2} 
  + |\Delta_{{\bf k},\lambda}^{\mathrm{s}}|^{2}
} 
\end{eqnarray}
for $\varepsilon_{{\bf k},\lambda}  > 0$, 
whereas for $\varepsilon_{{\bf k},\lambda}  \leq 0$ the $\pm$-signs have to be 
reversed. 

In order to find the renormalization equations (which governs 
the $\lambda$ dependence of the parameters of ${\cal H}_\lambda$), one has to 
evaluate the unitary transformation \eqref{3} explicitly. We use the 
following ansatz for the generator $X_{\lambda, \Delta \lambda}$
\begin{eqnarray}
\label{12}
 \lefteqn{X_{\lambda,\Delta \lambda} \,=\,}&& \\
 &=& \frac{1}{\sqrt{N}} \sum_{{\bf k},{\bf q} \in {\rm r.BZ}}
  \sum_{\gamma \in \{0,1\}} 
\left\{ b_{\gamma,{\bf q}}^{\dag} \underline{a}_{{\bf
  k},\lambda}^{\dag}\, \hat{A}_{\gamma,{\bf k},{\bf
    q}}^{\lambda,\Delta,\lambda} \,\underline{a}_{{\bf k} + {\bf q},\lambda} 
- \mbox{h.c.}   \right\} \nonumber
\end{eqnarray}
Besides the fact that the generator has to be anti-hermitian,  
$X_{\lambda,\Delta \lambda}^\dagger = - 
X_{\lambda,\Delta \lambda}$, the operator structure of the 
ansatz \eqref{12} agrees with that of the interaction ${\cal H}_{1,\lambda}$ 
of \eqref{7} where the electronic operators $\underline{c}_{\bf k}$ are 
replaced by the eigenmodes 
$\underline{a}_{{\bf k},\lambda}$ of ${\cal H}_{0,\lambda}$.
The matrix  $\hat{A}_{\gamma,{\bf k},{\bf q}}^{\lambda,\Delta,\lambda}$ 
in \eqref{12} has to
be fixed in such a way that, with respect to 
${\cal H}_{0,(\lambda - \Delta \lambda)}$, only excitations with energies 
smaller than $(\lambda-\Delta\lambda)$  contribute to 
$\mathcal{H}_{1,(\lambda- \Delta \lambda)}$. 
The renormalization equations 
for $\omega_{\gamma,{\bf q},\lambda}$ and 
the parameters of $\hat{H}^{\rm el}_{{\bf  k},\lambda}$ and 
$\hat{H}_{\gamma,{\bf k},{\bf q},\lambda}^{\rm w}$ 
are obtained by comparing Eqs.~\eqref{6} and \eqref{7} with the result of 
the explicit evaluation of \eqref{3} using ansatz \eqref{12} for 
$X_{\lambda, \Delta \lambda}$,
after all fermionic creation and annihilation operators
$\underline{a}_{\bf k}^{(\dagger)}$ 
have been transformed back to the original operators
$\underline{c}_{\bf k}^{(\dagger)}$. 
To evaluate Eq.~\eqref{3} an additional factorization 
approximation must be employed in order to keep only operators 
of the structure of those of \eqref{6} and \eqref{7}. 
Therefore, the final renormalization equations still depend on 
unknown expectation values. They are best evaluated
with the full Hamiltonian $\mathcal{H}$ in order to 
incorporate important interaction effects \cite{PRM}. Therefore, we have to
apply the sequence \eqref{3} of unitary transformations also 
to operators, 
$
  {\cal A}_{(\lambda-\Delta \lambda)}= e^{X_{\lambda,\Delta \lambda}} 
  {\cal A}_\lambda e^{-X_{\lambda, \Delta \lambda}}
$
and exploit
$
\langle{\cal A}\rangle = \langle {\cal A}_\lambda \rangle_{{\cal H}_\lambda} 
$.
This procedure is performed for the fermionic
and bosonic one-particle operators, $\underline{c}_{\bf k}^{(\dagger)}$ and
$b_{\gamma,{\bf q}}^{(\dag)}$, where the same approximations 
are used as for the Hamiltonian. The final set of coupled
renormalization equations is solved numerically.
Thereby, the equations for the expectation values are taken 
into account in a self-consistency loop.
The fully renormalized Hamiltonian is obtained for 
$\lambda \rightarrow 0$, where the interaction ${\cal H}_1$ 
is completely used up for the renormalization
of the parameters of ${\cal H}_{0,\lambda \rightarrow 0} 
=\tilde{\mathcal{H}}$. Thus,  an effectively free model is obtained,   
\begin{eqnarray}
\tilde{\mathcal{H}} &=& \sum_{{\bf k}\in {\rm r.BZ}} \underline{c}_{\bf
  k}^\dag \,\tilde{\hat{H}}^{\rm el}_{\bf
  k} \, \underline{c}_{\bf k} + \sum_{{\bf q} \in {\rm r.BZ}}
  \sum_{\gamma \in \{0,1\}}
\tilde{\omega}_{\gamma,{\bf q}} \, b_{\gamma,\bf q}^{\dag}
 b_{\gamma,\bf q} + \tilde{E}, \nonumber \\
\label{13}
&&
\end{eqnarray}
where
$
\tilde{\hat{H}}^{\rm el}_{\bf
  k}  = \hat{H}^{\rm el}_{{\bf k},\lambda \rightarrow 0} 
$
has the form as $\hat{H}_{{\bf k}, \lambda}^{\rm el}$ in  Eq.~\eqref{6}.
The final Hamiltonian $\tilde{\cal H}$ is diagonal and can be used
to investigate various physical properties. Note, in particular that the
eigenenergies $\tilde{E}_{1/2, {\bf k}}$ and 
$\tilde{\omega}_{\gamma, {\bf q}}$ of $\tilde{\cal H}$ 
can be considered as quasiparticles of the full Hamiltonian.    

Two aspects of the presented PRM approach should be noticed at this point: 
Even though a factorization approximation must be employed in order to 
derive the effectively free model \eqref{13}, fluctuations beyond 
mean-field theory are taken into account due to the renormalization procedure.
Furthermore, both electron-phonon interaction and phononic degrees of freedom 
are included in the microscopic model so that a direct access to phonon 
properties is provided.


\begin{figure}
  \begin{center}
    \scalebox{0.55}{
      \includegraphics*{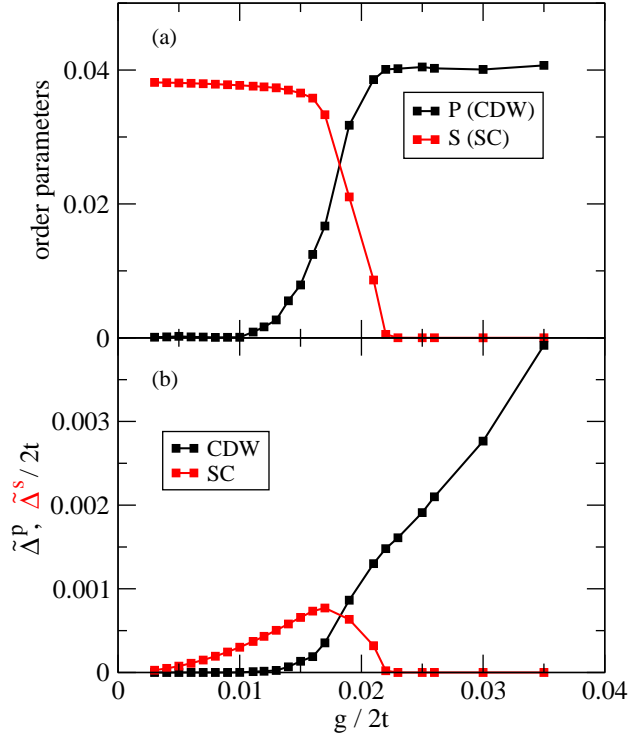}
    }
  \end{center}
  \caption{
    (Color online)
  (a) Order parameters $P = \frac{1}{N} \sum_{\bf k}
  \langle c_{{\bf k},\sigma}^{\dag} c_{{\bf k} 
  + {\bf Q},\sigma} \rangle$ (black line) and $S = \frac{1}{N} \sum_{\bf k}
  \langle c_{{\bf k},\uparrow}^{\dag} c_{-{\bf k},\downarrow}^{\dag} \rangle$
  (red line) as function of the electron-phonon coupling  
  $g$ for a square lattice with 144 lattice sites at half-filling,  
  $\omega_0 / t = 0.1$ and $T=0$. 
  (b) Renormalized values of the Peierls gap
  $\tilde{\Delta}_{\bf k}^{\rm p}$ (black line) and of the superconducting gap
  $\tilde{\Delta}_{\bf k}^{\rm s}$ (red line) at wave vector 
  ${\bf k} = (\pi / 2 , \pi / 2)$ for the
  same parameter values as in the upper panel.
  }
  \label{Fig_1/2}
\end{figure}

\begin{figure}
  \begin{center}
    \scalebox{0.55}{
      \includegraphics*{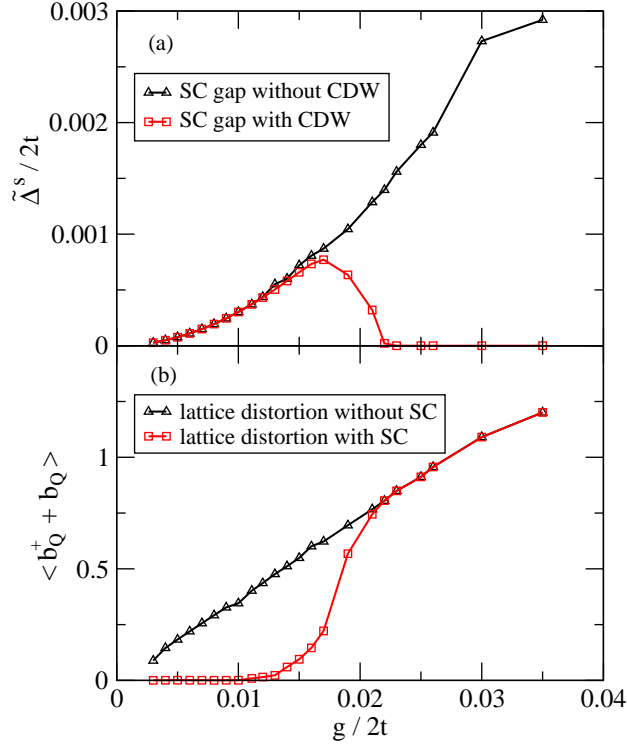}
    }
  \end{center}
  \caption{
    (Color online)
  (a) Renormalized superconducting gap as function of $g$
  for the case that CDW order is suppressed by hand (black curve).
  The red curve shows the complete solution including CDW order. 
  (b) Dimerized lattice distortion $\langle
  b_{\bf Q}^{\dag} + b_{\bf Q} \rangle$ as function of $g$. In the 
    black line a possible superconducting phase 
  is suppressed   The red line shows the complete solution 
  allowing also a superconducting state.
  }
  \label{Fig_3/4}
\end{figure}

\textit{Results and discussion}.
We consider a square lattice with $N=144$ sites 
and restrict ourselves to the Holstein model
with dispersion-less phonons $\omega_{\bf q} = \omega_0$. A local 
electron-phonon coupling  $g_{\bf q} = g$ is assumed. 
The temperature is set equal to $T=0$ 
and a small value of $\omega_0$ is chosen, 
$\omega_{0} = 0.1t$. In the following we concentrate on the $s$-wave like 
superconducting pairing because we could not find any stable $d$-wave solution.
Panel (a) of Fig.~\ref{Fig_1/2} shows the order parameters
for the Peierls state, ${P} = \frac{1}{N} \sum_{\bf k}
\langle c_{{\bf k},\sigma}^{\dag} c_{{\bf k} + {\bf Q},\sigma} \rangle$
(black), and for the superconducting state, 
${S} = \frac{1}{N} \sum_{\bf k}
\langle c_{{\bf k},\uparrow}^{\dag} c_{-{\bf k},\downarrow}^{\dag} \rangle$
(red) as function of $g$, where $g$ is
restricted to small values  $g/2t \leq 0.04$.
For comparison, in panel (b) of Fig.~\ref{Fig_1/2} the ${\bf k}$ dependent 
symmetry breaking fields $\tilde{\Delta}_{\bf k}^{\rm p}$ (black) and 
$\tilde{\Delta}_{\bf k}^{\rm s}$ (red) are shown for 
${\bf k} = (\pi / 2 , \pi / 2)$ for the same
parameters as in panel (a). 
Note that due to \eqref{11}, 
$\tilde{\Delta}_{\bf k}^{\rm p}$ and 
$\tilde{\Delta}_{\bf k}^{\rm s}$ contribute together to the energy gap 
in the quasiparticle spectrum of $\tilde{\cal H}$ and
either $P$ and $S$ or  
$\tilde{\Delta}_{\bf k}^{\rm p}$ and 
$\tilde{\Delta}_{\bf k}^{\rm s}$ can be considered as order parameters. 
As can be seen,  for small values of $g/2t < 0.010$ the system is in a pure
superconducting state, i.e.~no charge order
is present ($P=0$). The superconducting 
gap increases roughly proportional to $g^2$.
In the intermediate  $g$ range,  $0.010 < g/2t < 0.023$, a coexistence
of both order parameter $P$ and $S$ is found. Thus, the system 
is in a combined superconducting-charge ordered state. The 
$g$ dependence of   
$\tilde{\Delta}^s$ is no longer quadratic as in the small $g$ regime. 
Instead $\tilde{\Delta}^s$
reaches a maximum value and drops down to zero with increasing 
$g$. For  $g/2t > 0.023$ the superconducting phase is
completely suppressed and the system is in a pure 
charge ordered state.  

The mutual influence of the two order parameters is considered in 
Fig.~\ref{Fig_3/4}. First, in panel (a) of Fig.~\ref{Fig_3/4} the renormalized 
superconducting energy gap $\tilde{\Delta}^s$ is shown as function 
of $g$ for two cases: (i) the former result from Fig.~\ref{Fig_1/2} 
which follows from the full renormalization equations (in red), and 
(ii) the result when the 
charge order is suppressed in the renormalization equations 
'by hand' (in black).
The comparison shows that  superconductivity becomes 
strongly suppressed when the charge order is present for large $g$ values. 

In panel (b) of Fig.~\ref{Fig_3/4}, the dimerized lattice displacement 
$\langle b_{\bf Q}^{\dag} + b_{\bf Q} \rangle$
is shown as function of $g$ for the same parameter values 
as in Fig.~\ref{Fig_1/2}. Note that a dimer-like shift of 
the ionic equilibrium positions comes always along with the presence 
of a charge density wave. Thus, the expectation 
values $\langle b_{\bf Q}^{\dag} + b_{\bf Q} \rangle$
are nonzero in the Peierls phase and can be considered as an alternative
order parameter. The red curve in panel (b) of Fig.~\ref{Fig_3/4} shows 
 $\langle b_{\bf Q}^{\dag} + b_{\bf Q} \rangle$ as it follows from the solution
of the full renormalization equations. For comparison, 
in the black curve the superconducting phase  
is artificially suppressed. 
The comparison of the two curves 
shows that for weak coupling  $g/2t < 0.01$, 
when the superconducting state exists, the Peierls state is strongly
suppressed. However, for larger values of $g$ the superconducting 
order parameter vanishes 
and the red curve converges to the black curve. 
For the case, when the superconductivity is suppressed,
the ionic shift $\langle b_{\bf Q}^{\dag} + b_{\bf Q} \rangle$  increases 
almost linearly with $g$ and a Peierls state 
is found up to $g \rightarrow 0$. In this respect 
the two-dimensional Holstein model differs from the 
one-dimensional Holstein model, where a Peierls state exists
only above a critical electron-phonon coupling $g_c >0$.

\textit{Summary.} The developed approach to the two-dimensional Holstein 
model has two particular advantages: Fluctuations beyond mean-field theory 
are taken into account, and the appearance of superconductivity and 
CDW state is directly proven on the basis of their order parameters. Thus, 
the presented PRM approach overcomes important limitations of QMC studies 
\cite{QMC} and of the Bilbro-McMillan model \cite{BM}, and our results 
provide a reliable proof that indeed superconductivity and lattice 
distortion coexist in a certain parameter range of the electron-phonon 
coupling.


\textit{Acknowledgments.} We would like to acknowledge stimulating and 
enlightening discussions with P.B.~Chakaraborty. This work was supported by 
the DFG through the research program SFB 463.


\end{document}